\documentclass[aps,prb,twocolumn,superscriptaddress]{revtex4}
\usepackage{graphicx}

\begin{document}

\title{Temperature induced spin coherence dissipation in quantum dots}

\author{F. G. G. Hernandez}\email{felix@ifi.unicamp.br}
\altaffiliation{on leave from Instituto de F\'isica Gleb Wataghin
UNICAMP, SP, Brazil.} \affiliation{Experimentelle Physik II,
Technische Universit\"{a}t Dortmund, D-44221 Dortmund, Germany}
\author{A. Greilich} \affiliation{Experimentelle
Physik II, Technische Universit\"{a}t Dortmund, D-44221 Dortmund,
Germany}
\author{F. Brito}
\affiliation{IBM T. J. Watson Research Center, Yorktown Heights,
NY 10598, USA}
\author{M. Wiemann} \affiliation{Experimentelle
Physik II, Technische Universit\"{a}t Dortmund, D-44221 Dortmund,
Germany}
\author{D. R. Yakovlev}
\affiliation{Experimentelle Physik II, Technische Universit\"{a}t
Dortmund, D-44221 Dortmund, Germany}
\author{D. Reuter}
\affiliation{Angewandte Festk\"{o}rperphysik, Ruhr-Universit\"{a}t
Bochum, D-44780 Bochum, Germany}
\author{A. D. Wieck}
\affiliation{Angewandte Festk\"{o}rperphysik, Ruhr-Universit\"{a}t
Bochum, D-44780 Bochum, Germany}
\author{M. Bayer}
\affiliation{Experimentelle Physik II, Technische Universit\"{a}t
Dortmund, D-44221 Dortmund, Germany}

\date{\today}

\begin{abstract}
The temperature dependence of electron spin coherence in singly
negatively charged (In,Ga)As/GaAs quantum dots is studied by
time-resolved Faraday rotation. The decoherence time $T_2$ is
constant on a $\mu$s scale for temperatures below 20 K, for higher
temperatures it shows a surprisingly sharp drop into the
nanoseconds range. The decrease cannot be explained through
inelastic scattering with phonons, and may be related with elastic
scattering due to phonon-mediated fluctuations of the hyperfine
interaction.
\end{abstract}


\maketitle

Solid-state systems are interesting for implementation of quantum
information processing because they may provide controllable
qubits sufficiently protected from environment-induced
classicality.\cite{chuang,hanson} Specifically, in semiconductor
quantum dots (QDs) a qubit can be defined by the two-level system
of a confined electron spin,\cite{loss98} which currently attracts
great attention because of its long relaxation times. The spin
relaxation can be characterized by two times scales, the
longitudinal relaxation time $T_1$ limited by inelastic
scattering, and the transverse relaxation time $T_2$ (also called
decoherence time), for which limitations may arise also from
elastic scattering. The relation between these times is
non-trivial and is often summarized by the simple relation
$\left(T_2\right)^{-1} = 2 \left(T_1\right)^{-1} +
\left(T_2^{\prime}\right)^{-1}$, where $T_2^{\prime}$ is the pure
or elastic decoherence time.

For the $T_1$ time of a QD electron spin a number of
investigations exist, both from experiment and theory. Compared to
higher-dimensional systems, the $T_1$ times are very much enhanced
because the QD confinement protects the spin from the main
inelastic scattering mechanism: the electron spin coupling with
its orbital motion. In high magnetic fields $T_1$ has been shown
to persist over tens of milliseconds or even longer at cryogenic
temperatures,\cite{elzerman04,kroutvar04} in accord with
theoretical calculations.\cite{khaetskii00} Further, its
dependence on external parameters such as temperature and magnetic
field for neutral and charged quantum dots has been
studied.\cite{eble06,bernardot06,Lombez07,UrbaszekPRB07}

On the other hand, the information about the $T_2$ time is still
limited. Considering that inelastic scattering would be the only
channel for decoherence, $T_{2}$ may be as large as $2T_{1}$.
However, studies at cryogenic temperatures show $T_2$-times in the
microseconds range, showing that the elastic relaxation channel
due to hyperfine interaction plays the dominant role under these
conditions.\cite{petta05,greilich_science} Recently, several
calculations for $T_2$ times have been
reported.\cite{merkulov02,Erlingsson,khaetskii02,Woods02,SemenovPRL04,Coish07,WuPRB08,Woods08,CoishPRB08}

An important figure of merit of electron spin qubits is stability
under temperature changes. A temperature increase enhances the
lattice phonon occupation, so that decoherence mechanisms
involving phonons gain importance. Here we study the QD electron
spin coherence as function of temperature. We show that coherence
can be initiated by short laser pulses for temperatures up to
$\sim$ 100 K. The coherence time, however, is temperature
independent only up to 15 K, above it shows a sharp drop. From
model calculations we conclude that this sharp drop is not related
to spin-orbit coupling but arises from hyperfine interaction
fluctuations involving phonons.

Time-resolved Faraday rotation (FR) studies using a pump-probe
technique have been performed on an ensemble of singly negatively
charged (In,Ga)As/GaAs QDs (see Ref. \cite{greilich_prl} for
details). The sample was immersed in the variable temperature
insert of a superconductor magnet for fields $B$ aligned
perpendicular to the optical axis. For optical excitation a
mode-locked Ti:Sapphire laser was used, emitting pulses with
1.5-ps duration at a rate of 75.6 MHz (corresponding to
$T_{R}$=13.2 ns pulse separation) with a photon energy tuned to
the QD ground state optical transition. Using a laser pulse
picker, we were able to increment the laser repetition period
$T_{R}$.

Decoherence time measurements on QD ensembles are constrained by
dephasing due to inhomogeneities in the ensemble. The electron
spin dephasing time $T^{\star}_{2}$ has been found to be on the
order of 10 ns only.\cite{greilich_prl,dutt,braun} This fast
dephasing can be overcome by exciting with a train of laser pulses
which synchronizes precessional phase modes of electron spin
subsets in the ensemble.\cite{greilich_science,prb_felix} This
mode-locking produces constructive interference patterns in the FR
spectrum due to focusing of ensemble inhomogeneities. Consequently
it allows one to recover the dynamics of a single QD by filtering
out $T_{2}$ from a $T^{\star}_{2}$
measurement.\cite{greilich_science}

\begin{figure}[!h]
\includegraphics[width=0.8\columnwidth,keepaspectratio]{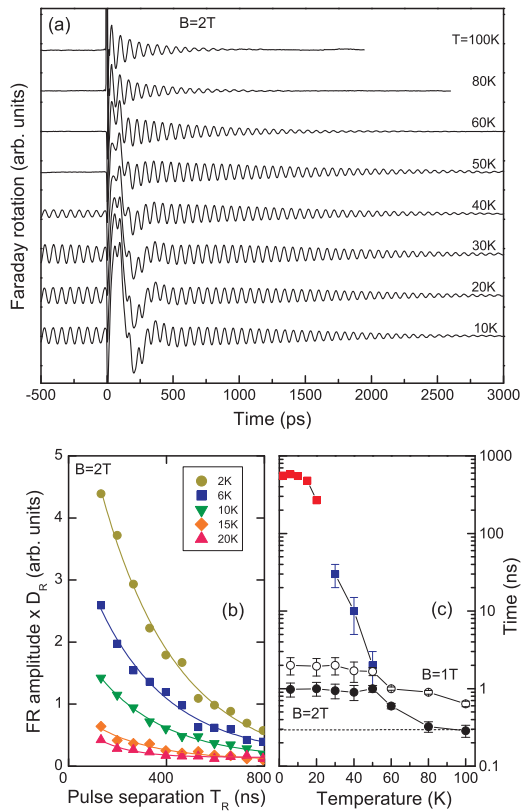}
\caption{(color online) (a) Normalized FR spectra vs pump-probe
delay at $B$ = 2 T for different temperatures. $P_{pump} =
180$~W$/$cm$^{2}$, $P_{probe} = 25$~W$/$cm$^{2}$. (b) FR amplitude
at negative delay vs laser repetition period $T_{R}$ for different
temperatures. Solid lines are fits using exponential decay forms
with time $T_{2}$. (c) Decoherence time $T_{2}$ (squares) and
dephasing time $T_{2}^{\star}$ (full circles) vs temperature at $B
= 2$ T. Open circles give $T_{2}^{\star}$ at $B = 1$ T. Dotted
line marks the exciton/trion lifetime.} \label{fig1}
\end{figure}

Figure \ref{fig1}(a) shows FR traces at $B = 2$ T for different
temperatures as function of delay between pump and probe. After
initialization of a spin pure state at time zero, coherent
oscillations due to spin precession about the magnetic field are
observed. Within the first ns of delay, the ensemble signal arises
either from resident electrons in singly charged QDs or from
exciton electrons in neutral QDs. The exciton lifetime is about
300 ps, as measured by differential transmission spectroscopy.
Therefore the FR signal after $\sim$ 1 ns can be related to
resident electrons only.\cite{greilich_prl}

At temperatures $T <$ 30 K the resident electron signal at
positive delays is accompanied by coherent signal on the negative
delay side. This signal arises from mode-locking of electron spins
whose precession frequencies are synchronized with the exciting
laser. When the temperature is increased above 40 K, the negative
delay signal disappears rather abruptly, while the positive delay
signal is still pronounced up to 100 K.\cite{comment} Thus the
data show that resident electron spin coherence can be efficiently
created at elevated temperatures. For an electron with arbitrary
spin the excitation creates a superposition of an electron state
that blocks excitation due to Pauli principle and a charged
exciton consisting of a spin singlet electron doublet and a hole.
After decay of the trion, an electron is left whose polarization
along the optical axis has been increased by the excitation. This
mechanism works, however, only if the hole spin is not scattered
during the pump pulse.\cite{greilich_prl}

For discussing spin coherence we focus on the negative delay
signal. As reported earlier,\cite{greilich_science} the spin
decoherence time $T_2$ may be inferred by measuring the FR
amplitude at negative delay for increasing separation $T_{R}$
between the laser pump pulses. A change in $T_{R}$ can be
expressed in terms of the division rate $D_R = T_R /$(13.2 ns). In
our studies $T_{R}$ was changed from 132.0 ns ($D_R$=10) up to
794.4 ns ($D_R$=60). The upper $D_R$ limit is set by the negative
delay signal becoming too weak due to the low cycling rate.

In figure \ref{fig1}(b) the negative delay FR amplitude at $B$ = 2
T, multiplied by $D_R$ to correct the spectroscopic response for
decreasing average power at fixed signal recording time, is
plotted vs the laser repetition period for different temperatures.
The experimental data are fitted by exponential decays with times
$T_{2}$ (solid lines).\cite{greilich_science} $T_{2}$  as function
of temperature is plotted in Fig.\ref{fig1}(c) by the squares. At
low temperature the measured spin coherence time $T_{2}$ is about
600~ns, in good accord with previous
reports.\cite{petta05,greilich_science} $T_{2}$ remains constant
with temperature increment up to 15 K. However, we find a
surprisingly sharp drop of $T_{2}$ down to 250~ns  at 20 K.

Heating up further, the negative delay FR signal can be seen only
for small pump laser separations, but a systematic increase of
$DR$, as required for measuring $T_2$, is not possible. E. g.,
strong mode-locking signal is seen at $T$ = 30 K for $T_R =
13.2$~ns, but for $D_R = 10$ the signal becomes already
unmeasurably small. This clearly suggests that the spin coherence
is destroyed on time scales far below 132~ns. In this temperature
range ($T > 30$~K ) we have therefore used the mode-locking
amplitudes at negative delay for $T_R = 13.2$~ns to obtain
estimates for $T_2$ (blue squares in Fig.\ref{fig1}(c)).
Calculations show that the coherence time cannot exceed 30 $\pm$
10 ns in order to loose the mode-locking signal completely when
increasing $D_R$ = from 1 to 10 at $T = 30$~K. The mode-locking
amplitude for $D_R = 1$ decreases strongly going from 30 to 40~K,
which can be explained by a further reduction of $T_2$ to 10
$\pm$5~ns. At $T$ = 50~K the mode-locking signal has vanished
completely for all $D_R$, which can be explained by a drop of
$T_2$ into the 2~ns range.

The decay times of the FR signal for positive delays are also
shown in Fig.\ref{fig1}(c) by the circles for $B$ = 1 and 2 T.
These times have been determined by fitting the FR traces by
exponentially damped harmonics with damping time $T_2^{\star}$. At
low temperatures the decay is determined by dephasing due to
ensemble inhomogeneities such as electron g-factor variations or
nuclear spin fluctuations. The relation between $T_2^{\star}$ and
$T_2$ is given by $\left(T_2^{\star}\right)^{-1} =
\left(T_2\right)^{-1} + \left(T_{inh}\right)^{-1}$, where the
second term is the inhomogeneity-related scattering rate.

For $T < 30$~K the dephasing time is basically constant and
exceeds 1 ns for the chosen experimental conditions. As $T_2$ is
by more than two orders of magnitude longer in this range,
$T_2^{\star}$ is basically identical to $T_{inh}$. While for $B <
0.5$~T the nuclear field fluctuations become important, for higher
fields the g-factor variations dominate. These variations are
translated into a precession frequency variation scaling linearly
with $B$, so that the dephasing occurs faster at 2 T than at 1 T
(see Fig.\ref{fig1}(c)). For completeness we note that under
mode-locking conditions the dephasing depends on optical pump
power. We use large bars to indicate this variation and not the
experimental error (see Ref. \cite{greilich_science} for details).
Above 30~K we find a drop of $T_2^{\star}$, which we attribute to
the increased importance of the homogeneous relaxation channel
$1/T_2$. From extrapolating the $T_2$ data one expects that $T_2$
becomes shorter than $T_2^{\star}$ for $T > 50$~K.

As mentioned, the main sources of electron spin decoherence are the
spin-orbit coupling and the hyperfine interaction. As for the first
case, because of the coupling of the orbital electronic motion to
acoustic phonons, the spin-orbit interaction leads to an indirect
dissipative channel. The spin-orbit coupling comprises two
interaction mechanisms due to bulk inversion asymmetry of the
crystal lattice (Dresselhaus) and asymmetry of the QD confining
potential (Rashba).\cite{hanson,TsitsishviliPRB04,Bulaev} Both
decoherence contributions can be investigated by mapping the
interaction Hamiltonians onto bath-of-oscillators models in which
the spin is directly coupled to the
bath.\cite{caldeiraleggett,harry03} The corresponding spin-boson
Hamiltonian is given by:
\begin{equation}
\hat{H}_{eff}=-\frac{\hbar}{2}\Delta\hat{\sigma}_x+\sum_i\hbar\omega_i\hat{b}_i^\dagger\hat{b}_i+\hat{\sigma}_z\sum_i
c_i\left(\hat{b}_i^\dagger+\hat{b}_i\right),\label{hamiltonian}
\end{equation}
where $\Delta=g\mu_BB/\hbar$ is the Zeeman frequency with electron
g-factor $g$, $\omega_i$ is the phonon frequency, and
$\hat{\sigma}$ is the Pauli matrix. The second and third terms
describe the oscillator bath. Here
$\{\hat{b}_i,\hat{b}_i^\dagger\}$ are bosonic annihilation and
creation operators. The third term account for the spin-bath
coupling.

The details of the Dresselhaus and Rashba interactions are
comprised in the effective spectral function $J_{eff}(\omega)$ of
the bath ``seen'' by the electron spin. If the applied magnetic
field $B$ is such that the Zeeman frequency $\Delta$ is much less
than the bath resonance peak $\Omega$ ($\Delta/\Omega\ll1$), the
spin dissipative dynamics occurs in the low frequency regime of
the effective spectral function, given by:\cite{harry03}
\begin{equation}
J_{eff}(\omega)\approx
m^\ast\gamma^2\delta_s\left(\frac{\omega_D}{\omega_0}\right)\left(\frac{\omega}{\omega_D}\right)^{s+2},\label{jeff}
\end{equation}
where $m^\ast$ is the electron effective mass and $\omega_0$ is
the splitting between the confined electron states, $\delta_s$ is
the dimensionless electron phonon coupling, and $\omega_D$ is the
Debye frequency. The parameter $\gamma$ corresponds to the
spin-orbit coupling $\beta$ and $\alpha$ for the Dresselhaus and
Rashba contributions, respectively. The exponent $s$ distinguishes
piezoelectric ($s=3$) and deformation potential ($s=5$)
interaction.

In the spin-bath weak coupling limit, the Bloch-Redfield equations
\cite{Hartmann} can be used to determine the spin expectation
values, $\sigma_i={\rm Tr}\rho\hat{\sigma}_i$. Solving these
equations \cite{brito}, we find $T_1$ and $T_2$ times related by
$T_2=2T_1$ (in agreement with Ref.\cite{Golovach}):
\begin{equation}
\frac{1}{T_{2,SO}}=\frac{1}{4}J_{eff}(\Delta)\coth\left(\frac{\hbar\Delta}{2k_B
T}\right).\label{T2}
\end{equation}

\begin{figure}[!h]
\includegraphics[width=0.8\columnwidth,keepaspectratio]{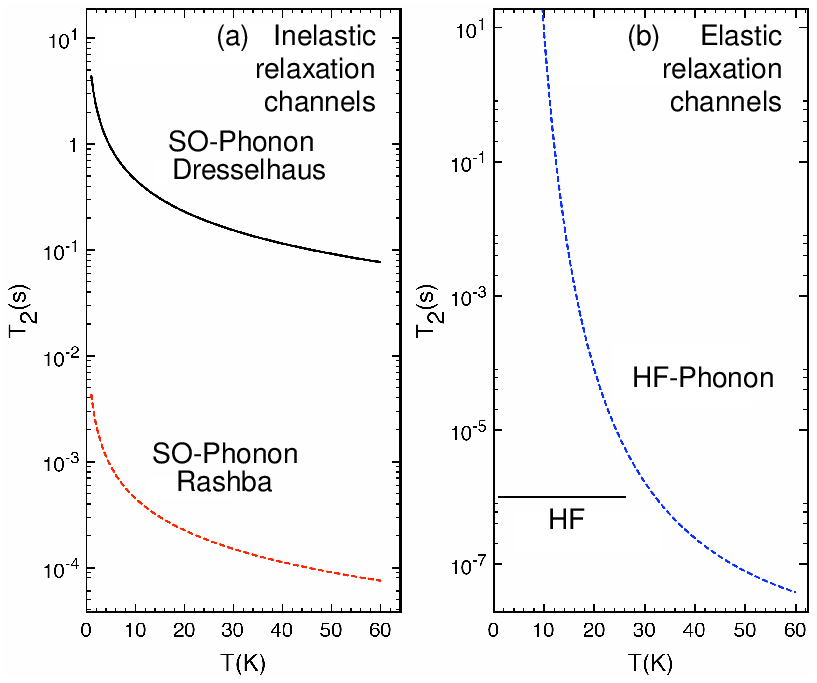}
\caption{Calculated temperature dependence of $T_2$ for
In$_{0.5}$Ga$_{0.5}$As QDs. $T_2$ due to (a) inelastic
spin-orbit-phonon scattering and (b) elastic scattering time for
hyperfine interaction (solid horizontal line \cite{WuPRB08}). In
the right panel, the dashed line includes hyperfine interaction
fluctuations due to phonon involvement.\cite{SemenovPRL04}}
\label{fig2}
\end{figure}

The left panel of Fig.~\ref{fig2} shows the calculated decoherence
times $T_2$ as function of temperature up to 60~K, assuming
piezoelectric interaction for the Dresselhaus and Rashba
interactions.\cite{comment2} We assumed an In$_{0.5}$Ga$_{0.5}$As
QD composition with an electron level splitting
$\hbar\omega_0=20$~meV and a Zeeman energy
$\hbar\Delta=69.5~\mu$eV (for example, corresponding to $B=2$~T,
$g=0.6$), as for the experimentally studied QDs. In addition, the
following parameter values were used: $m^\ast=0.041m_e$,
$\delta_3=298.5$, $\omega_D=27.5$~meV, and the spin-orbit
couplings $\beta=3\times 10^3$~m/s (Dresselhaus) and
$\alpha=9.6\times 10^4$~m/s
(Rashba).\cite{TsitsishviliPRB04,landoldt} We note that the
calculated relaxation rates vary with In-composition only by
factors of order unity. For both interactions we see in the panel
(a) a strong drop of the transverse spin relaxation time. From the
comparison we see that the Rashba interaction is three orders of
magnitude more efficient than the Dresselhaus interaction. Still
over the whole range the calculated times are orders of magnitude
longer than the experimentally observed $T_2$. Therefore we
exclude spin-orbit coupling as source for the observed spin
decoherence.

This leaves us with the hyperfine interaction described by a
Hamiltonian which couples the electron spin ${\bf S}$ and the
$i$-th nuclear spin ${\bf I}_i$ in the QD:
\begin{eqnarray}
\hat{H}_{HF} = \sum_i A_i \mid \psi \left( \bf{R}_i \right) \mid ^2
\left( \hat{S}_z \hat{I}_{i,z} + \hat{S}_+ \hat{I}_{i,-} +
\hat{S}_- \hat{I}_{i,+} \right),
\end{eqnarray}
where the sum goes over all nuclei in the QD electron localization
volume. The interaction strength is determined by the hyperfine
constant $A_i$ and the electron density $\mid \psi \left( \bf{R}_i
\right) \mid ^2$ at the nuclear site $\bf{R}_i$. $\hat{H}_{HF}$ mediates
processes in which the spins of electron and nucleus are mutually
flipped, as described by the products of raising and lowering
operators $\hat{S}_\pm$ and $\hat{I}_{i,\pm}$, which increase and
decrease the spin projections $S_z$ and $I_{i,z}$ along the
quantization axis $z$, respectively.

Indications of an inelastic scattering channel have been found in
studies of the dynamic nuclear polarization (DNP) by interaction
with an optically oriented electron \cite{UrbaszekPRB07}. The DNP
was found to be moderately increased for temperatures $<$ 50 K.
This was attributed to a temperature induced increase of the spin
flip-flop efficiency by phonon induced broadening of the electron
level. This efficiency is restricted at cryogenic temperatures
because of the mismatch in energy splittings between the electron
and the nuclear Zeeman levels. The phonons required for
compensating the energy mismatch are ``frozen'' under these
conditions. By a temperature induced level broadening this energy
mismatch may be softened. The data in Ref.\cite{UrbaszekPRB07},
however, suggest that the change of the inelastic scattering by
$\sim$10\% is too weak to explain the strong drop observed
experimentally.

Independent of inelastic scattering, calculations of elastic
mechanisms involving the hyperfine interaction have found
decoherence times in the $\mu$s-range \cite{khaetskii02,Woods02}, as
indicated in Fig.\ref{fig2}(b) by the horizontal line taken from
Ref.\cite{WuPRB08}. Recently, theoretical calculations proposed an
efficient decoherence mechanism due to modulations of the hyperfine
field by phonons that may be dominant at low magnetic fields and
high temperatures.\cite{SemenovPRL04} The corresponding decoherence
time can be estimated by:
\begin{equation}
\frac{1}{T_{2,HF}}=\Gamma\left(I_i,n_{Ii},V_{QD},A_i\right)
F\left(\frac{\hbar\omega_{0}}{2k_B T}\right),\label{T2hf}
\end{equation}
where $\Gamma$ is a function of the nuclear spins concentration
$n_{Ii}$ in the QD volume $V_{QD}$; and $F(x)=(1-\tanh^{2}x)\tanh
x$ contains the temperature dependence.

Equation \ref{T2hf} in combination with our QD parameters
($\Gamma^{-1}\sim$ 2.89~ns) is plotted in Fig.~\ref{fig2}(b) by
the dashed curve. The results agree with the $T_2$ drop observed
experimentally at about the same temperatures. The calculation
deviation from the data can be related with the difficulty to
determine the precise QD composition and the resulting nuclear
environment of the electron spin.

In conclusion, we observed that the temperature induced
decoherence time dependence in (In,Ga)As self-assembled QDs shows
two regimes: (i) $T < 20$ K: $T_{2}$ is temperature independent
and limited by the hyperfine interaction, (ii) $T > 20$~K: $T_{2}$
is strongly temperature dependent and the main driven decoherence
mechanism may be related with phonon-mediated hyperfine
interaction fluctuations. One can see from Eq.~(5) describing the
resulting decoherence time, $T_2$ may be stabilized towards higher
$T$ by increasing the level splitting $\omega_0$ of the QDs.

This work was supported by the BMBF-project nanoquit, the DFG (BA
1549/12-1) and the FAPESP (contract 04/02814-6).


\begin{references}
\bibitem{comment}
We also observe spin coherence for temperatures higher than 100 K
but then the dephasing times drop into the range of the trion
lifetime. Several reasons may contribute: Besides fast hole
scattering, electron scattering can be relevant as LO phonons become
available, which may even lead to thermal emission of electrons so
that the coherent signal may increasingly be contributed by neutral
excitons. As no clear distinction is possible at the moment, we
restrict to temperatures below 100 K.
\end{references}
\end{document}